\documentclass[10pt,a4paper]{article}

\usepackage{graphicx}
\usepackage{dcolumn}
\usepackage{bm}

\newcommand{\de}{\partial}

\newcommand{\eps}{\epsilon}

\newcommand{\etal}{{\em et al.}}
\newcommand{\etals}{{\em et al.\ }}

\newcommand{\cc}{\textrm{c.c.}}

\begin{document}

\title{All-optical diode action with quasiperiodic photonic crystals}
\author{Fabio Biancalana \\ School of Physics and Astronomy, Cardiff
University \\ CF24 3AA, Cardiff (UK)}

\maketitle

\begin{abstract}
We theoretically investigate the possibility of realizing a
nonlinear all-optical diode by using the unique features of
quasiperiodic 1D photonic crystals. The interplay between the
intrinsic spatial asymmetry in odd-order Thue-Morse lattices and
Kerr nonlinearity, combined with the unconventional field
localization properties of this class of quasiperiodic sequences,
gives rise to sharp resonances that can be used to give a
polarization-insensitive, nonreciprocal propagation with a contrast
close to unity for low optical intensities.
\end{abstract}


\section{Introduction and motivations}

An all-optical diode (AOD) is a spatially nonreciprocal device that
in the ideal case, and for a specific wavelength $\lambda$, allows
the total transmission of light along the forward direction
($T^{+}_{\lambda}=1$), and totally inhibits light propagation along
the backward direction ($T^{-}_{\lambda}=0$), yielding a unitary
contrast
$C=(T^{+}_{\lambda}-T^{-}_{\lambda})/(T^{+}_{\lambda}+T^{-}_{\lambda})=1$.
AODs are widely considered to be the key components for the next
generation of all-optical signal processing, in complete analogy
with electronic diodes which are widely used in computers for the
processing of electric signals. Replacing relatively slow electrons
with photons as carriers of information would substantially increase
the speed and the bandwidth of telecommunication systems, leading to
a real revolution of the telecom industry. However, controlling
photons is a much more difficult task than controlling electrons,
due to the fact that the latter particles are electrically charged,
and thus much more suitable for obtaining the nonlinear
characteristics typical of the electric diode.

Due to the crucial technological implications, such unidirectional
propagation ('diode action') has been studied experimentally by
several groups, using the most diverse schemes and experimental
techniques. Photonic crystals (PCs) \cite{yablo}, i.e. multilayered
structures which possess a definite periodicity of the same order of
the wavelength of light, are very suitable for this kind of
applications because of their ability to totally inhibit the
propagation of light over well-defined frequency regions, known as
photonic bandgaps (PBGs, see Refs. \cite{yablo,yeh}). Especially
noticeable in this direction are the early efforts of Scalora \etals
\cite{scalora}, who in 1994 proposed the use of 1D PCs, with a
gradation that makes the structure spatially asymmetric, allowing
nonreciprocity due to different forward/backward nonlinear shifts of
the band edges in the transmission spectrum. More recently, in 1999
Gallo and Assanto \cite{gallo1} proposed the use of a waveguide with
an asymmetrically placed defect, which breaks the forward/backward
symmetry, in presence of quadratic nonlinearities. Most of the above
schemes, however, suffer from some serious drawbacks which make them
not suitable for commercial and large-scale applications. Relatively
large physical sizes are often needed; the balance between figures
of merit and optical intensities is usually inadequate, and in some
cases cumbersome structural designs are necessary to provide
structural asymmetry. For instance, in the widely used optical
isolators, which allow a strong unidirectional propagation and are
based on the Faraday rotation effect \cite{faraday}, the use of two
polarizers is necessary, because the effect is strongly
polarization-dependent, which makes the device very efficient but a
few centimeter long.

In this work, we discuss a different type of AOD design based on 1D
nonlinear multilayered structures arranged in a quasiperiodic or
aperiodic fashion \cite{albuquerque}. Since the first report by
Schechtman \etals \cite{firstquasi} of metallic Al-Mn alloys showing
a quasiperiodic crystal structure, and the pioneering experimental
works of Merlin \etals on Fibonacci and Thue-Morse (T-M) GaAs-AlAs
superlattices \cite{merlin}, a large number of studies have been
devoted to 1D, 2D and 3D quasicrystals, see \cite{albuquerque} and
references therein. More specifically, we examine the diode action
in T-M lattices, which exhibit isolated, sharp and high quality
resonances inside the so-called pseudobandgaps, i.e. bandgap regions
which are not due to translational symmetry. Thanks to the intrinsic
asymmetry of odd-order T-M sequences, and due to the unique field
localization properties of T-M sequences, we have found that it is
possible to achieve nonreciprocal nonlinear shifts of these
'pseudoresonances' by using small optical intensities, which
combined with the small size of the structures (of the order of a
few microns), would make them feasible for advanced applications and
for integration on chip.

\section{Linear transmission of T-M photonic crystals}

A quasiperiodic chain is defined by a recursive, deterministic
generation rule $S_{0}=\{A\}$, $A\rightarrow f_{1}(A,B)$ and
$B\rightarrow f_{2}(A,B)$, where $S_{0}$ is the initial string and
$f_{1,2}(A,B)$ are two arbitrary strings of symbols $A$ and $B$. $A$
and $B$ indicate the two kinds of layer of which the structure is
made of, with refractive indices $n_{A}$ and $n_{B}$ respectively.
Generation $S_{j}$ is a string which is constructed by applying the
above recursive rule $j$ times. Several basic quasiperiodic chains
have been extensively investigated in the literature, such as the
Fibonacci, Cantor, Double-Period and Rudin-Shapiro sequences, see
also \cite{albuquerque} and references therein. The quasiperiodic
structure that we consider here, and one of the most important in
general, is the T-M sequence, which was introduced by Thue in a 1906
study \cite{thue}. It is based on the rule $\mathcal{T}_{0}=\{A\}$,
$A\rightarrow\{AB\}$, $B\rightarrow\{BA\}$. The size of the string
$\mathcal{T}_{j}$ for the T-M sequence is given by $2^{j}$, so that
for real-life applications one usually considers $j<10$. The first
T-M sequences are: $\mathcal{T}_{1}=\{AB\}$,
$\mathcal{T}_{2}=\{ABBA\}$, $\mathcal{T}_{3}=\{ABBABAAB\}$, etc. The
ratio $L_{j}$ between number of blocks of type $A$ and number of
blocks of type $B$ in $\mathcal{T}_{j}$ is always equal to unity,
irrespective of the generation number $j$. It is important to note
that the T-M sequence $\mathcal{T}_{j}$ is symmetric with respect to
string inversion operation $\mathcal{I}$ (which is performed by
reversing the order of the string elements) if $j$ is an even
integer, while it is antisymmetric (i.e. it is invariant under
$\mathcal{I}$ plus the substitution $A\rightarrow B$ and
$B\rightarrow A$) if $j$ is an odd integer. In the present work we
consider only antisymmetric sequences, in that asymmetry, together
with nonlinearity, is an essential requisite for the nonreciprocal
behavior which is at heart of the optical diode action.

Figs. 1(a,b,c,d) show the linear transmission spectra for plane
waves (as calculated with the transfer matrix method, TMM, see Ref.
\cite{yeh}) for the T-M photonic crystals $\mathcal{T}_{3,5,7,9}$
respectively, in the wavelength range $0.5$-$1.1$ $\mu$m. Normal
incidence is assumed, i.e. vanishing incidence angle $\theta=0$, so
that TE- and TM-polarization spectra are degenerate. At this stage,
forward and backward spectra also coincide
($T^{+}_{\lambda}=T^{-}_{\lambda}$, $\forall\lambda$), since the
linear transfer matrix $M$ is unimodular \cite{yeh}, and, despite
the strong asymmetry of the considered lattices, no optical diode
action is possible in the linear regime. Crystal parameters are:
refractive indices of the layers $n_{A}=1.55$ (which corresponds to
the polydiacetylene $9$-BCMU organic material, see Refs.
\cite{organic,scalora}), $n_{B}=2.3$ (which corresponds to TiO$_{2}$
material), $\Delta n\equiv n_{B}-n_{A}=0.75$, layer thicknesses
$d_{A}=\lambda_{0}/4/n_{A}\simeq 112.9$ nm and
$d_{B}=\lambda_{0}/4/n_{B}\simeq 76.1$ nm, where $\lambda_{0}\simeq
0.7$ $\mu$m (indicated with a dashed red line in Fig. 1). Total
physical lengths of the crystals are approximately $0.756$ $\mu$m
for $\mathcal{T}_{3}$ [$8$ layers], $3.02$ $\mu$m for
$\mathcal{T}_{5}$ [$32$ layers], $12.09$ $\mu$m for
$\mathcal{T}_{7}$ [$128$ layers] and $48.38$ $\mu$m for
$\mathcal{T}_{9}$ [$512$ layers].

Several things can be deduced from the panel of Fig. 1. First of
all, it is easy to see that, when increasing the generation number
$j$, each peak of transmission shows evidence of self-similarity.
This is a very well-known feature which is discussed in many works
\cite{fractal}. Secondly, most of the peaks shown in Fig. 1 are {\em
peaks of perfect transmission} (PPTs, for which $T=1$), regardless
of the value of $j$, and they tend to accumulate around the
reference frequency $\lambda_{0}$, at which the quarter wavelength
condition is satisfied. We have found that, amongst the basic
quasiperiodic crystals, this unexpected property seems to be a
unique and remarkable feature of T-M sequences. Another important
feature shown in Fig. 1 is the gradual (for $j\geq 7$) formation of
{\em pseudobandgaps}, i.e. frequency regions of zero transmission
(not due to periodicity) where the crystal behaves like a perfect
mirror. In Figs. 1(c) and 1(d) one particular pseudobandgap under
scrutiny is indicated with a green shaded region, while other
pseudobandgaps (the ones containing other PRs) present in the
spectra are indicated with blue areas. This gap contains a sharp
resonance [{\em pseudoresonance}, PR, indicated with blue arrows in
Fig. 1(c) and 1(d)], analogous to a defect state, that in presence
of nonlinearity can be used to obtain nonreciprocal propagation.
This is located at $\lambda_{res}\simeq 0.8643$ $\mu$m. The
generation of the above sharp resonances can be attributed to the
specific localization phenomena in T-M sequences
\cite{localization}. Both extended and critical states exist in T-M
sequences, which give rise to two distinct types of bandgaps. One is
the traditional bandgap that exists in periodic crystals, the other
one shows fractal (self-similar) splitting into smaller bands. The
above types of gaps are distinguished by their different behavior
when increasing the generation number \cite{localization}.
Traditional gaps change their depths only, while fractal gaps split,
creating the sharp peaks of perfect transmission which are
impossible to obtain with periodic crystals or resonators.

Quality factor (Q-factor) is a measure of the average energy in the
resonant peak over the energy radiated per cycle \cite{qfactor}. It
is defined by $Q\equiv \lambda_{res}/\Delta\lambda$, where
$\lambda_{res}$ is the central resonant wavelength and
$\Delta\lambda$ is the resonance full-width at half maximum (FWHM).
In Fig. 2(a) the evolution of Q-factor for the PR indicated above is
shown as a function of generation number. The Q-factors for the
PR$_{j}$ resonance are: $Q(PR_{6})=262.06$, $Q(PR_{7})=864.8$,
$Q(PR_{8})=1736.53$ and $Q(PR_{9})=1235.4$. The resonance appears
(together with the pseudobandgap) at $j=6$ [Fig. 2(b)], and it has a
well-defined bell-shape until $j=9$ [Fig. 2(e)], at which the
fractal self-replication of the spectrum occurs. At this point the
PR broadens due to the birth of satellite peaks. Remarkably, the
central wavelength of the PR (where it exists) does not depend on
$j$. For the chosen index contrast, it thus makes sense to talk
about this PR only for $j=6,7,8$. However, we need to restrict
ourselves to odd-order T-M sequences, the only ones that do not
exhibit spatial symmetry. For this reason we select
$\mathcal{T}_{7}$ as our representative structure, which has the
best PR specifics for the selected value of $\Delta n$. In order for
the pulse to clearly distinguish the resonance, its duration
$\tau_{p}$ must be $\gg 2\pi/\Delta\omega$, where $\Delta\omega$ is
the resonance FWHM. In the case of PR$_{7}$ we have
$\Delta\omega\simeq 2.5$ THz, so that a quasi-CW pulse is defined by
$\tau_{p}>2.5$ psec. This, together with material relaxation time,
will limit the device switching time.

Before embarking in the nonlinear analysis of the structures, it is
useful to understand how small random fluctuations in the angle of
incidence of light from perfect normal incidence ($\theta=0$) affect
the transmission spectra for TE- and TM-polarizations. Fig. 3 shows
the $\theta$-dependence of TE- and TM-polarization spectra for
$\mathcal{T}_{7}$. Position of the PR$_{7}$ peak is also indicated.
For $\theta=0$ (normal incidence), the two output spectra must
coincide. The most important feature that can be deduced from Fig. 3
is the presence of minima in the peak positions in correspondence of
the 'degeneracy' point $\theta=0$, a desirable feature for what
concerns device applications. This leads to a solid {\em structural
stability} of the PR to small random fluctuations in $\theta$, which
will therefore be polarization-independent to a high degree of
approximation.

\section{Nonlinear transmission and optical diode action}

We now proceed to analyze the nonlinear behavior of the PR$_{7}$
peak shown in Fig. 2(c). In order to do this, we use the nonlinear
TMM method outlined in Ref. \cite{scalora}. According to this
method, for each value of the input wavelength $\lambda$, we must
guess the output forward amplitude, while the backward output
amplitude vanishes. After propagating back to the input layer, one
obtains a value for the total input intensity, which is proportional
to the sum squared of the input forward and backward fields. We
repeat this process until the input intensity matches the desired
incident intensity. Figs. 4(a,b) display a sketch of the different
nonlinear frequency shifts of PR$_{7}$ for three different input
intensity conditions ($I=0$, black dotted lines; $I=10$ MW/cm$^{2}$,
solid red lines; and $I=50$ MW/cm$^{2}$, solid blue lines), for both
forward [Fig. 4(a), $T^{+}_{\lambda}$] and backward [Fig. 4(b),
$T^{-}_{\lambda}$] incidence. The nonlinear Kerr refractive indices
for TiO$_{2}$ and $9$-BCMU used in our calculations are roughly
estimated to be $n_{2}(\textrm{TiO}_{2})\simeq 10^{-14}$ cm$^{2}$/W
and $n_{2}(\textrm{$9$-BCMU})\simeq 2.5\cdot 10^{-11}$ cm$^{2}$/W.
These are measured values \cite{data,organic}, obtained by using
degenerate and non-degenerate pump-probe experiments. Results
published in \cite{organic} show that linear and nonlinear
absorptions can be minimized considerably in $9$-BCMU inside certain
spectral regions, which are therefore suitable for device operation.
Thus here we neglect these two absorptions for simplicity. The
relaxation time $\tau_{rel}$ of $9$-BCMU is estimated to be around
$1.6$ psec. The switching time of the device is fast and equal to
$\max\{\tau_{p},\tau_{rel}\}$.

Let us take, for instance, an input pump intensity equal to
$I=I_{p}=10$ MW/cm$^{2}$, corresponding to the curves displayed with
red solid lines in Figs. 4(a,b). Note the strong difference in
transmission between forward and backward incidence. Moreover, there
is a reduction of the transmission peaks with increasing intensity.
By tuning the pump wavelength near the maximum of the forward
incidence peak [$\lambda_{p}\simeq 0.8659$ $\mu$m, see green
vertical dashed line in Fig. 4(a)], one has substantial transmission
for forward incidence of pump light ($T_{\lambda_{p}}^{+}=0.905$).
However, for the same wavelength, the backward transmission is
located on the tail [see green vertical dashed line in Fig. 4(b)],
which ensures a very low transmission for the backward direction
($T_{\lambda_{p}}^{-}=0.06$). For the particular case shown in Figs.
4(a,b), this gives an efficient diode action with an AOD contrast
equal to $C=0.875$. The resonant nature of the PR enhances the
nonlinear effects to a point that the AOD action is obtained by
using modest intensities.

\section{Pulse propagation}

Finally, in order to confirm the results obtained above for
realistic quasi-CW pulses, which have a small but finite spectral
width well inside the resonance, we accurately solve Maxwell's
equations in presence of an inhomogeneous permittivity $\eps(z)$.
Let us write the 1D Maxwell equations in the conventional form
$\de_{t}[\eps E]+c\de_{z}B=0$, $\de_{t}B+c\de_{z}E=0$, where
$E(z,t)$ and $B(z,t)$ are the scalar electric and magnetic fields
respectively, $c$ is the speed of light in vacuum.
$\eps(z,t)=\eps_{0}(z)+\eps_{2}I$ is the dielectric permittivity,
where $\eps_{0}(z)^{1/2}=n(z)$ is the refractive index (real and
$\geq 1$), and $\eps_{2}$ is the Kerr-nonlinear dielectric constant.
$I\equiv|E(z,t)|^{2}$ is the electric field intensity. We then
decouple the fast spatiotemporal oscillations due to the carrier
wavenumber $k_{0}$ and frequency $\omega_{0}$ with the expansion:
$E(z,t)=[A(z,t)e^{ik_{0}z-i\omega_{0}t}+\cc]/2$,
$B(z,t)=[G(z,t)e^{ik_{0}z-i\omega_{0}t}+\cc]/2$. $A$ and $G$ are the
complex envelopes of the electric and magnetic fields respectively,
and $k_{0}=\omega_{0}/c$. After defining new dimensionless variables
$\xi\equiv z/z_{0}$ and $\tau\equiv t/t_{0}$, where $z_{0}$ and
$t_{0}$ are respectively arbitrary spatial and temporal scales, we
choose $z_{0}=c t_{0}$ and $t_{0}=2\pi/\omega_{0}$. We also rescale
the fields as $\{A,G\}\rightarrow A_{0}\{A,G\}$, where $A_{0}$ is a
convenient reference intensity. This leads to the dimensionless
equations
\begin{eqnarray}
\de_{\tau}A+\frac{1}{\eps}\de_{\xi}G+2\pi i(G/\eps-A)=0, \label{main1}\\
\de_{\tau}G+\de_{\xi}A+2\pi i(A-G)=0, \label{main2}
\end{eqnarray} which are suitable for numerical computations, and
{\em do not rely on any slow-varying amplitude approximation}. An
important observation is that the nonlinear dielectric function
$\eps$ is given in the new variables by
$\eps(\xi,\tau)=\eps_{0}(\xi)+(3/4)\eps_{2}A_{0}^{2}|A(\xi,\tau)|^{2}$,
a non-trivial expression that satisfies simultaneously the terms
rotating with the first and the third harmonics.
Eqs.(\ref{main1}-\ref{main2}) are solved by first initializing $A$
as a function of $\xi$, at $\tau=0$, and then posing
$G(\xi,\tau=0)=\pm n(\xi)A(\xi,\tau=0)$, where the plus (minus) sign
refers to a forward (backward) propagation condition, which is used
to test the forward/backward asymmetry in our T-M structure.

Results of simulations are shown in Fig. 4(c,d). Fig. 4(c) shows the
nonlinear transmission of a forward supergaussian pulse with a
temporal width of $\tau_{p}=2$ psec and a spatial width
$\xi_{p}=600$ $\mu$m ($\gg$ than the structure's length, which is
$12.09$ $\mu$m), and input wavelength $\lambda_{p}=0.8659$ $\mu$m,
corresponding to the location of the green dashed line in Figs.
4(a,b). The chosen supergaussian profile well approximates the
incidence of a finite CW wave, and input intensity was taken to be
$I=10$ MW/cm$^{2}$. According to the previous discussions, this
allows almost unitary transmission ($T^{+}\approx 0.9$) if the
spectral width of the pulse is narrower than that of the PR.
Launching the same pulse in the backward direction we have a
dramatically different behavior: Fig. 4(d) shows the nonlinear
transmission in this case, where light is almost entirely reflected
($T^{-}\approx 0.05$), confirming both qualitatively and
quantitatively the optical diode action and the plane-wave
predictions discussed above.

\section{Conclusions}

In conclusion, we have studied the linear and nonlinear transmission
properties of a micron-size T-M quasiperiodic PC near a PR, i.e. a
sharp resonance located in the pseudobandgap. Sharp resonances are
due to the unique field localization properties of T-M sequences. We
have shown that the strong asymmetry of odd-order Thue-Morse
lattices, combined with a Kerr nonlinearity, gives rise to a highly
nonreciprocal transmission which is the major feature of an AOD.
This effect has also been confirmed by numerically integrating
Maxwell's equations for realistic optical pulses. Our design allows
an unprecedented reduction in size of the device at relatively low
operational optical intensities, a consequence of the intrinsic
antisymmetry of the structures considered, and the resonant,
strongly localized nature of the transmission.

The author would like to thank Victor Grigoriev for useful comments.
This work is supported by the UK Engineering and Physical Sciences
Research Council (EPSRC).

\newpage

{\bf Fig. 1} Linear transmission spectra for plane waves at normal
incidence ($\theta=0$) for asymmetric odd-order T-M photonic
crystals (a) $\mathcal{T}_{3}$, (b) $\mathcal{T}_{5}$, (c)
$\mathcal{T}_{7}$, (d) $\mathcal{T}_{9}$. Reference wavelength
$\lambda_{0}=0.7$ $\mu$m is indicated with a red dashed line. The
pseudobandgap and the PR of interest for this paper are indicated
with a green area and a blue arrow respectively. Other large
pseudobandgaps which contain other PRs are indicated with blue
areas. Crystal parameters are: $n_{A}=1.55$, $n_{B}=2.3$,
$d_{A}=\lambda_{0}/4/n_{A}\simeq 112.9$ nm and
$d_{B}=\lambda_{0}/4/n_{B}\simeq 76.1$ nm.

\hspace{5mm}

{\bf Fig. 2} (a) Q-factors of PR at $\lambda_{res}=0.8643$ $\mu$m as
a function of generation number for T-M PCs. (b,c,d,e) Transmission
spectra of the $\lambda_{res}=0.8643$ $\mu$m PR for $j=6,7,8,9$
respectively. Crystal parameters are the same as in Fig. 1.

\hspace{5mm}

{\bf Fig. 3} $\theta$-dependence of TM- and TE-polarization spectra
(left and right parts of the plot respectively) for the {\em linear}
T-M quasicrystal $\mathcal{T}_{7}$. Large blue regions correspond to
pseudobandgaps. Crystal parameters are the same as in Fig. 1.

\hspace{5mm}

{\bf Fig. 4} (a,b) Sketch of nonlinear shifts for forward [(a)] and
backward [(b)] incidence, for three different input intensities.
(c,d) Results of pulsed simulations, showing the asymmetric
transmission properties of the nonlinear device for forward [(c)]
and backward [(d)] incidence. In both cases, input intensity is
$I=10$ MW/cm$^{2}$. The input pulse has a supergaussian profile,
$A(\xi,\tau=0)=\sqrt{I}\exp[-(\xi/\xi_{p})^{m}]$, with $m=4$ and
$\xi_{p}=600$ $\mu$m, which corresponds to a pulse duration of
$\tau_{p}=2$ psec. The position of the T-M crystal is indicated by a
vertical bar. Crystal parameters are the same as in Fig. 1.

\newpage

\begin{figure}
\includegraphics[width=\columnwidth]{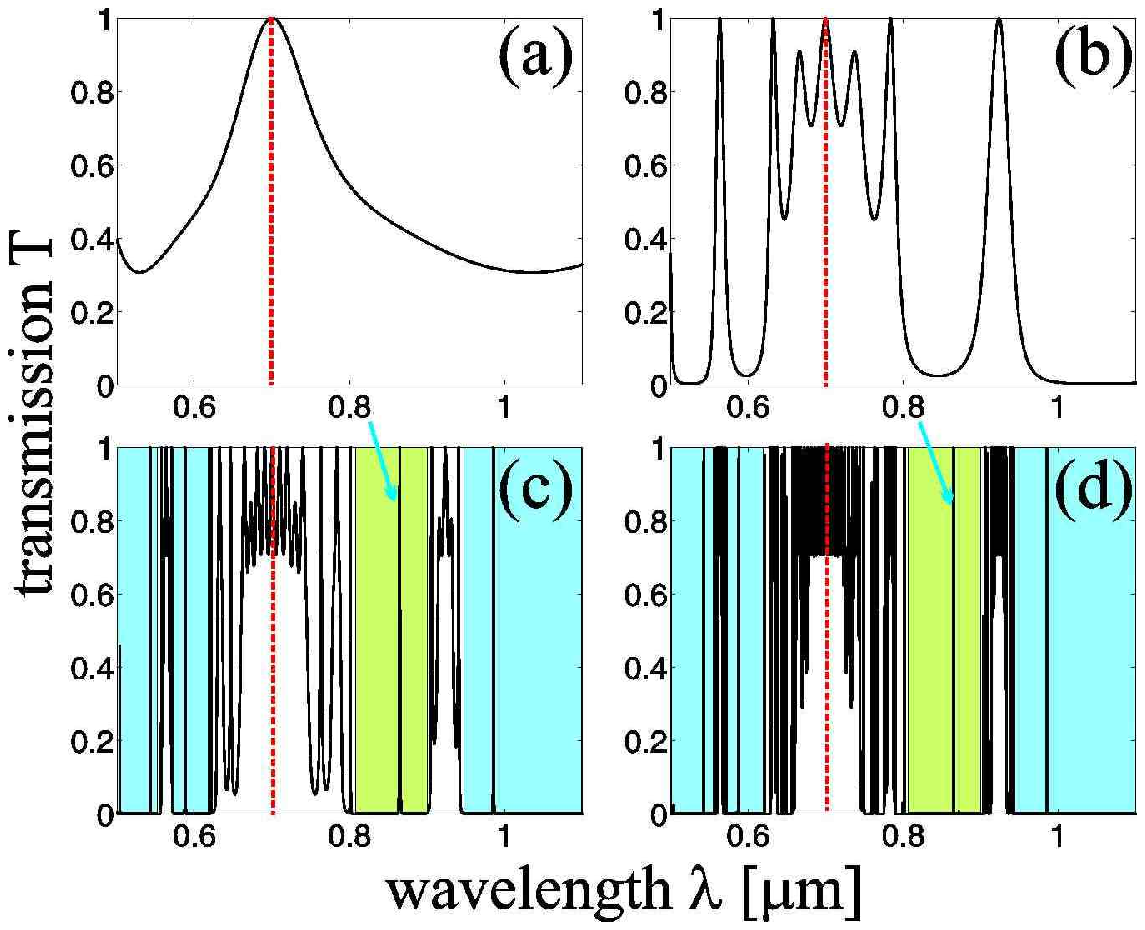}
\caption{}
\end{figure}

\begin{figure}
\includegraphics[width=\columnwidth]{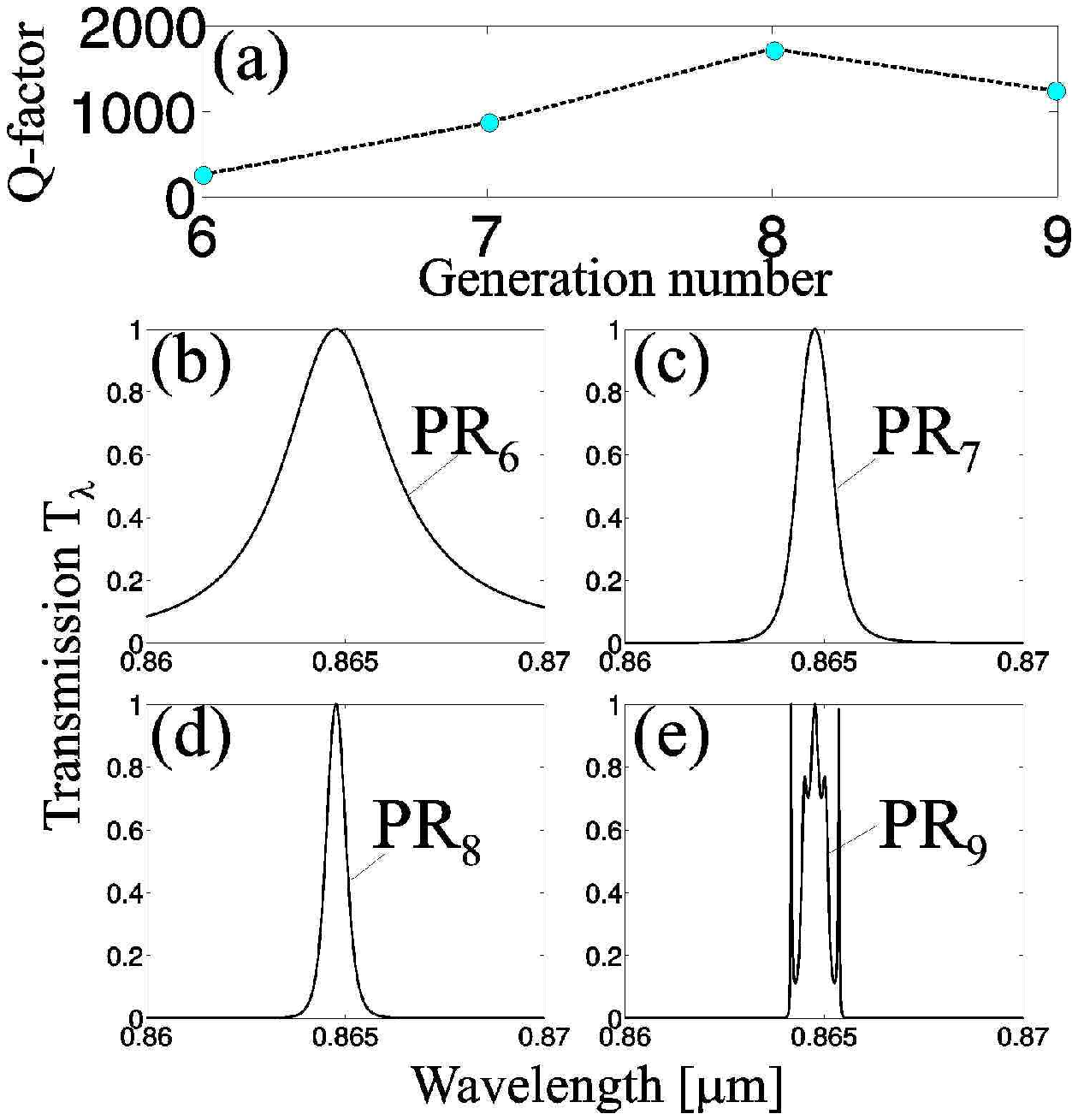}
\caption{}
\end{figure}

\begin{figure}
\includegraphics[width=\columnwidth]{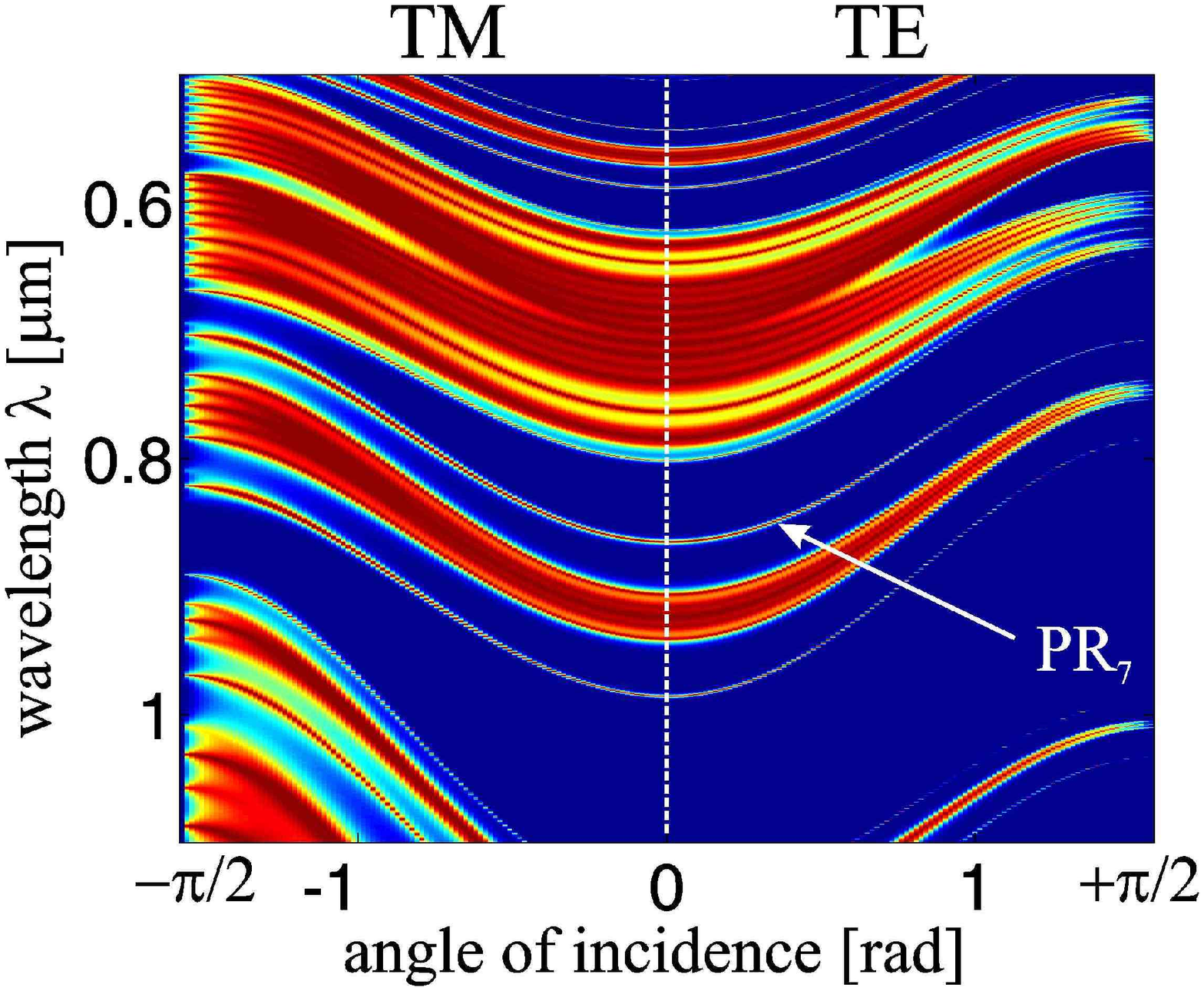}
\caption{}
\end{figure}

\begin{figure}
\includegraphics[width=\columnwidth]{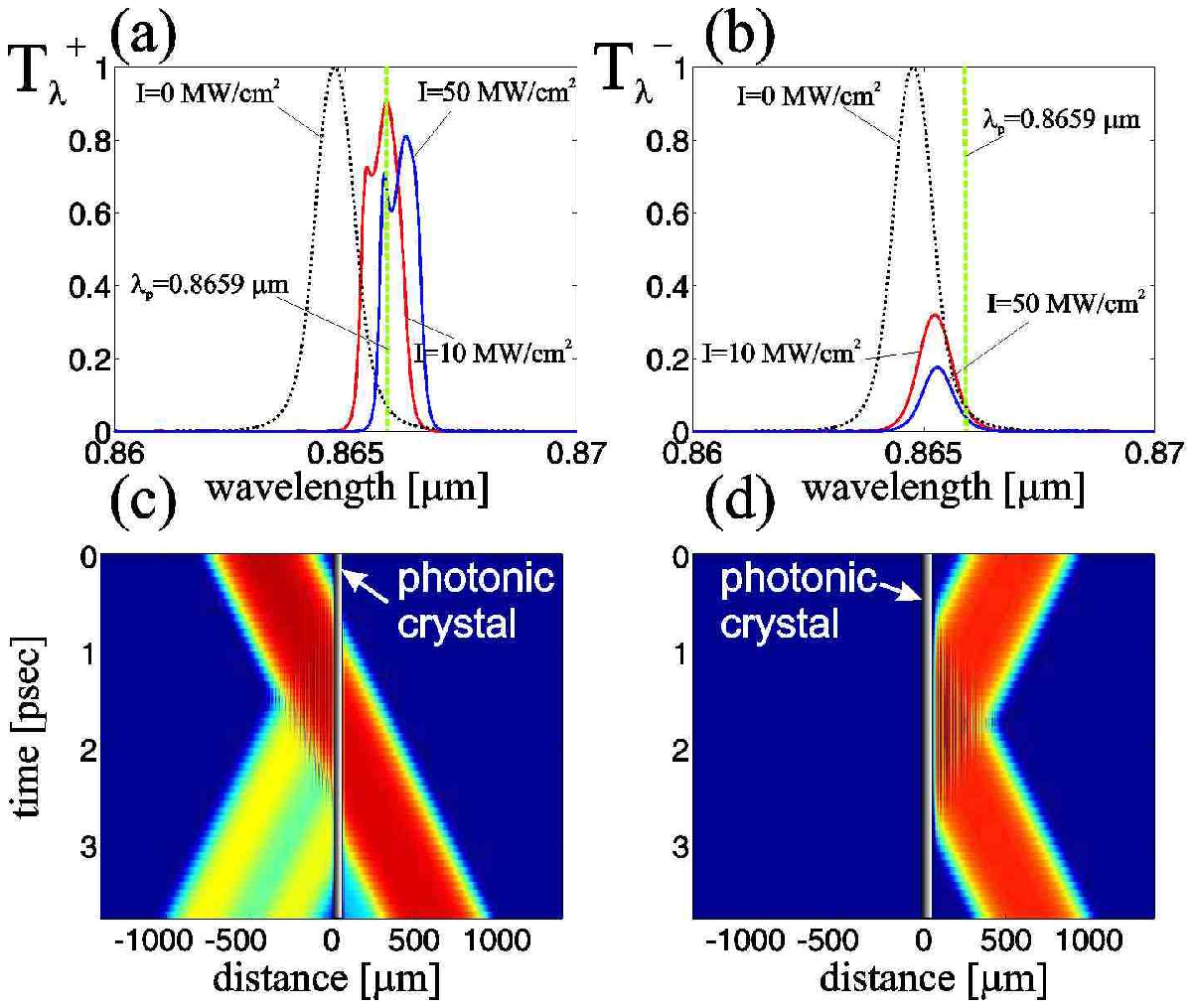}
\caption{}
\end{figure}


\newpage

\begin{thebibliography}{99}

\bibitem{yablo} E. Yablonovitch, Phys. Rev. Lett. {\bf 58}, 2059
(1987); S. John, Phys. Rev. Lett. {\bf 58}, 2486 (1987).

\bibitem{yeh} P. Yeh, {\em Optical Waves in Layered Media} (Wiley, New Jersey,
2005).

\bibitem{scalora} M. Scalora \etal, J. Appl. Phys. {\bf 76}, 2023 (1994); Tocci
\etal, Appl. Phys. Lett. {\bf 66}, 2324 (1995).

\bibitem{gallo1} K. Gallo and G. Assanto, Opt. Lett. {\bf 16}, 267
(1999); K. Gallo \etal, Appl. Phys. Lett. {\bf 79}, 314 (2001).

\bibitem{faraday} L. J. Aplet and J. W. Carlson, Appl. Opt. {\bf 3},
544 (1964).

\bibitem{albuquerque} E. L. Albuquerque and M. G. Cottam, {\em Polaritons in Periodic and Quasiperiodic
Structures} (Elsevier, Amsterdam, 2004).

\bibitem{firstquasi} D. Shechtman \etal, Phys. Rev. Lett. {\bf 53},
1951 (1984).

\bibitem{merlin} R. Merlin \etal, Phys. Rev. Lett. {\bf 55}, 1768
(1985); Z. Cheng, R.Savit and R. Merlin, Phys. Rev. B {\bf 37}, 4375
(1988).

\bibitem{thue} A. Thue, Norske Vididensk. Selsk. Skr. I. {\bf 7},
1 (1906).

\bibitem{organic} S. Molyneux \etal, Opt. Lett. {\bf 18}, 2093
(1993); A. K. Kar, Polym. Adv. Technol. {\bf 11}, 553 (2000).

\bibitem{fractal} Nian-hua Liu, Phys. Rev. B {\bf 55}, 3543
(1997).

\bibitem{localization} W. Gellermann \etal, Phys.
Rev. Lett. {\bf 72}, 633 (1994); H. Hiramoto and M. Kohmoto, Phys.
Rev. Lett. {\bf 62}, 2714 (1989); V. Agarwal \etal, Photon.
Nanostruct. {\bf 3}, 155 (2005); X. Y. Jiang \etal, Appl. Phys.
Lett. {\bf 86}, 201110 (2005); J. M. Luck \etal, J. Phys. A: Math.
Gen. {\bf 26}, 1951 (1993).

\bibitem{qfactor} D. Ripin \etal, IEEE J. Light. Tech. {\bf 17},
2152 (1999).

\bibitem{data} R. Adair, L. L. Chase and S. A. Payne, Phys. Rev. B {\bf
39}, 3337 (1989).

\end{thebibliography}
\end{document}